\documentclass[twocolumn,prb,aps,showpacs,superscriptaddress]{revtex4}
\usepackage{graphicx}% Include figure files

\begin{document}

\title{Proximity Effect in Gold Coated $YBa_2Ca_3O_{7-\delta}$  Films Studied by Scanning Tunneling Spectroscopy}

\author{Amos Sharoni}
\affiliation{Racah Institute of Physics, The Hebrew University,
Jerusalem 91904, Israel}
\author{Itay Asulin}
\affiliation{Racah Institute of Physics, The Hebrew University,
Jerusalem 91904, Israel}
\author{Gad Koren}
\affiliation{Department of Physics, Technion - Israel Institute of
Technology, Haifa 32000, Israel}
\author{Oded Millo}
\email{milode@vms.huji.ac.il} \affiliation{Racah Institute of
Physics, The Hebrew University, Jerusalem 91904, Israel}

\begin{abstract}
Scanning tunneling spectroscopy on gold layers over-coating
\textit{c}-axis $YBa_2Ca_3O_{7-\delta}$ (YBCO) films reveals
proximity induced gap structures. The gap size reduced
exponentially with distance from \textit{a}-axis facets,
indicating that the proximity effect is primarily due to the (100)
YBCO facets. The penetration depth of superconductivity into the
gold is $\sim 30$ nm, in good agreement with estimations for the
dirty limit. The extrapolated gap at the interface is  $\sim 15$
meV, consistent with recent point-contact experiments.  The
proximity-induced order parameter appears to have predominant
\textit{s}-wave symmetry.
\end{abstract}


\maketitle

The mutual influence of a superconductor (S) in good electrical
contact with a normal metal (N), a phenomenon known as the
proximity effect (PE), has been studied extensively for
conventional low-temperature superconductors.\cite{1,2}  However,
the picture of the PE is not as clear for the high-temperature
\textit{d}-wave superconductors, such as $YBa_2Ca_3O_{7-\delta}$
(YBCO), both experimentally and theoretically.  In particular,
local probe measurements of the PE, such as those performed for
conventional proximity systems,\cite{3,4} are still lacking for
these systems. In addition to the fundamental interest, the issue
of the proximity effect is significant also from the point of view
of applications, \textit{e.g.}, for superconductive electronic
devices based on the Josephson effect.

In a homogeneous conventional (\textit{s}-wave) superconductor,
the gap in the density of states (DOS) corresponds to the
superconducting pair-potential (PP), a measure of the ability of
quasi-particles to form Cooper pairs.\cite{5} In an N-S proximity
structure, where the PP is not spatially uniform, the gap in the
DOS does not necessarily reflect the PP. Ideally, an abrupt change
in the pair potential can take place at the N-S interface: from a
finite PP on the S side, to zero on the N side. However, the gap
in the DOS, which is a measure of the local pair amplitude, may
change smoothly across the interface, from the full bulk value
deep in the S side to zero at a distance characterized by the
penetration depth into N.\cite{1,3,4,6,7}

For an anisotropic \textit{d}-wave superconductor, the
crystallographic orientation of the superconductor surface at the
N-S interface can modify significantly the PE.  For example,
Josephson coupling was observed in high temperature S-N-S
junctions when the normal to the junction plane was along the
\textit{a}-axis (transport into the a-b plane), but not when it
was along the \textit{c}-axis (transport perpendicular to the a-b
plane).\cite{8,9} This implies that Cooper pairs can leak into N
only along the $CuO_2$ planes.  The penetration depth (the normal
coherence length), $\xi_n$, of the Cooper pairs into N was
extracted from the temperature and N-thickness dependencies of the
critical current.\cite{9,10,11} However, a direct local probe
measurement showing the way in which the pair amplitude decays in
N has not yet been performed. Another unresolved issue is the
symmetry of the proximity-induced order parameter in N, when S has
a bulk \textit{d}-wave PP symmetry. Theoretical calculations show
that the order parameter induced in the N side can have an
\textit{s}-wave symmetry.\cite{12,13} Kohen \textit{et al.} found
a $d_{x^{2}-y^{2}} - is$ PP at Au-YBCO point contact junctions,
where the s component showed a systematic enhancement with
increased junction transparency,\cite{14} reaching a value of
around 16 meV. This behavior was attributed to an unusual
proximity effect that modifies the PP in YBCO near the N-S
interface. In particular, it induces an \textit{s}-wave component
accompanied by a reduction in the dominant  $d_{x^{2}-y^{2}}$-wave
PP.  A question arises now regarding the connection between this
\textit{s}-wave component and the proximity-induced order
parameter at the N side of the interface.

The PE can be directly investigated by measuring the evolution of
the proximity-induced gap in the normal-metal DOS as function of
the distance from the N-S interface, using scanning tunneling
microscopy (STM).\cite{3,4}  The magnitude of this gap is a
measure of the local proximity-induced pair amplitude. In this
work, we used \textit{c}-axis YBCO films covered with gold layers
of different thickness to study this gap evolution, in correlation
with the gold thickness and the local surface morphology.
(Unfortunately, this geometry does not allow monitoring the PE
properties at the S side, as previously done by Levi \textit{et
al.} for $Cu-NbTi$ junctions.\cite{3}) In the present study, the
tunneling spectra revealed an exponential decay of the proximity
gap with distance from \textit{a}-axis facets. This indicates that
the PE is primarily due to the interface between the normal metal
and the (100) YBCO surface, and not with the (001) surface.
Interestingly, while this facet-selectivity reflects the in-plane
versus out of plane anisotropy in YBCO, the tunneling spectra
measured on the Au layer exhibited isotropic behavior, suggesting
an \textit{s}-wave proximity-induced order parameter.  The
superconductor penetration depth into the gold film was found to
be $\xi_n \sim 30$ nm, in good agreement with theoretical
estimations for the dirty limit, and the extrapolated gap at the
interface was about 15 meV, in accordance with the above
observation of Kohen \textit{et al.}.\cite{14}

\begin{figure}
\includegraphics[width=8cm]{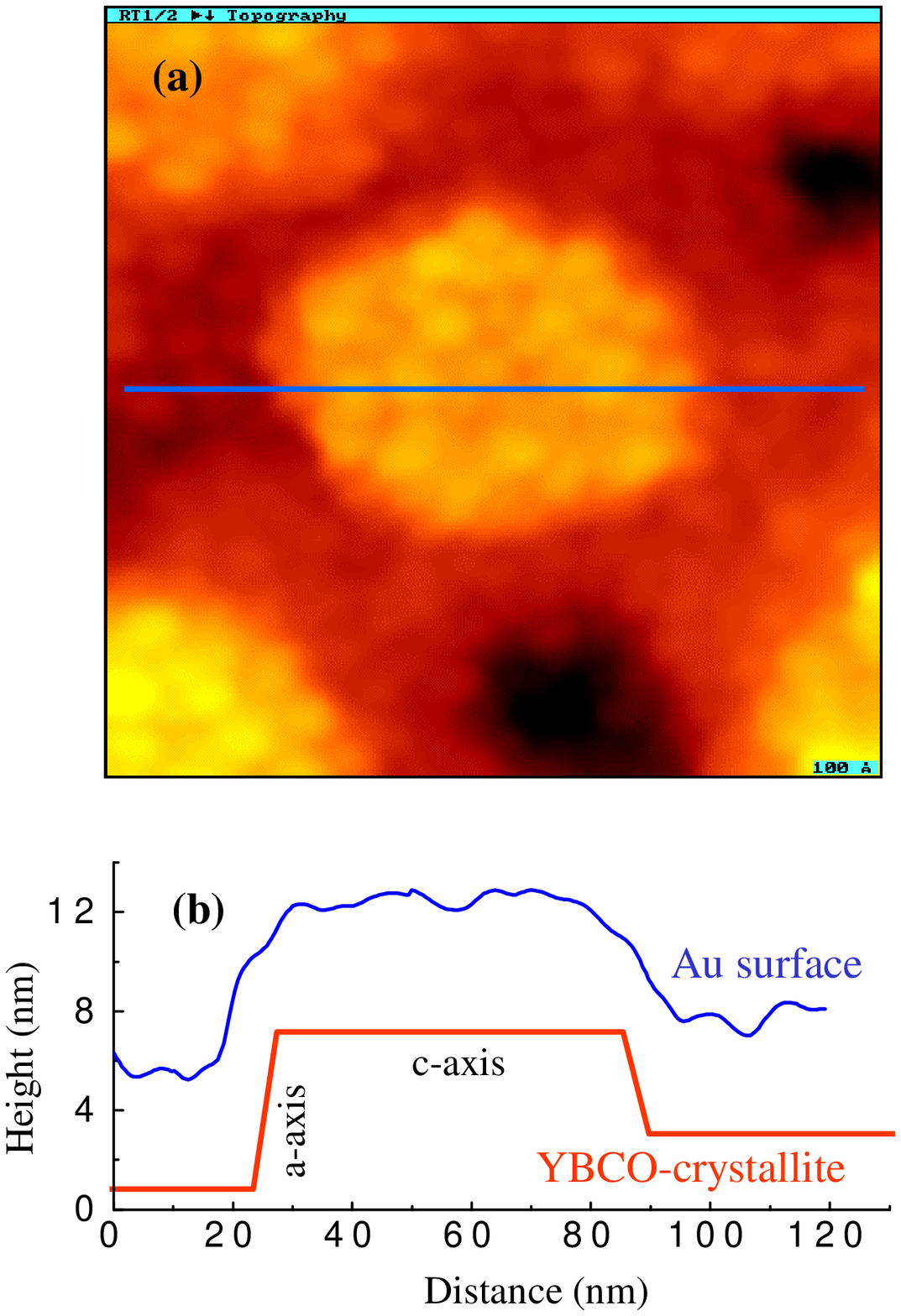}
\caption{(a) Topographic STM image ($100\times100$ nm$^{2}$) of an
YBCO film coated by a 5 nm thick Au layer, showing an YBCO
crystallite with clear facets and the granular morphology of the
Au film. (b) A cross-section measured along the line drawn in (a)
is shown together with a schematic cross-section of the underlying
YBCO crystallite, with labeled facets.}\label{fig1}
\end{figure}

Optimally doped epitaxial YBCO films of 50 nm thickness were grown
on (100) $SrTiO_3$ wafers by laser ablation deposition, with
\textit{c}-axis orientation normal to the substrate, as described
elsewhere.\cite{15} Next, a gold layer was deposited in-situ by
laser ablation at a temperature of 150 $^{0}$C  at  a rate of 1
\AA /s, and annealed for 1 hour at this temperature before cooling
down to room temperature. A total of 12 samples were prepared with
gold thickness varying between 1.5 nm to 60 nm. A bare reference
YBCO film showed a sharp (0.5 K wide) transition at 90 K and gap
values of 18 to 20 meV. The samples were transferred from the
growth chamber in a dry atmosphere, and introduced into our
cryogenic STM after being exposed to ambient air for less than 15
minutes. The YBCO films consisted of nearly square-shaped
crystallites, 50-100 nm lateral size and 10-15 nm
height,\cite{15,16} which had relatively large (100) facets. This
crystalline structure was clearly observed, yet somewhat smeared,
even after deposition of the thickest Au layer. The gold films
revealed a granular morphology with surface grains of $\sim$10 nm
in lateral size and rms height roughness of up to 1.5 nm (see
Fig.\ref{fig1}).

Tunneling spectra (dI/dV \textit{vs.} V characteristics) were
obtained by numerical differentiation of I-V curves that were
measured in correlation with the topography by momentarily
disconnecting the feedback loop. About 10 curves were acquired at
each position to assure data reproducibility.  We also checked the
dependence of the tunneling spectra on the voltage and current
settings (\textit{i.e.}, the tip-sample distance, or the tunneling
resistance values - in the range of 100 $M\Omega$  to 1 $G\Omega$)
and found that it does not affect the measured gap features. This
rules out the possibility that these gaps are even partially
related to single electron charging effects.\cite{17}

\begin{figure}
\includegraphics[width=8cm]{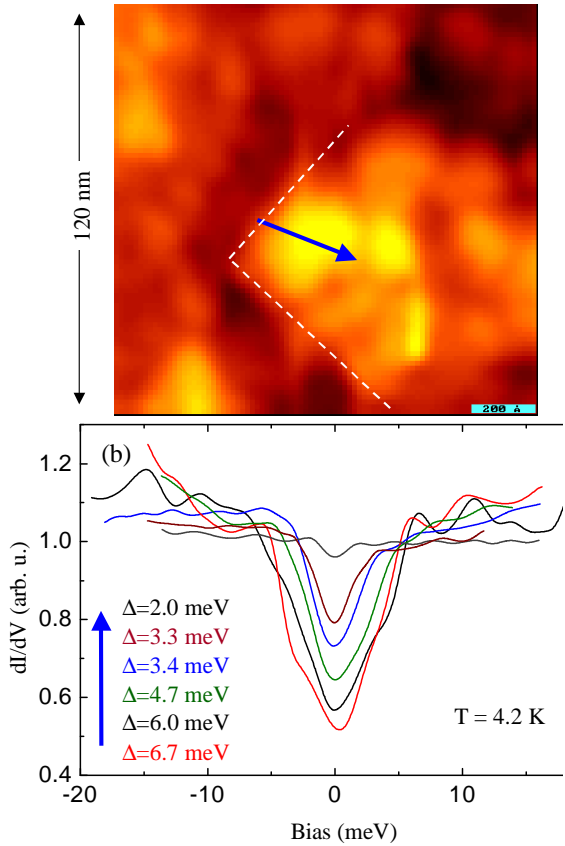}
\caption{A typical measurement showing the decrease in gap size as
a function of the distance from a (100) facet at a crystallite
edge. (a) Topographic image showing an YBCO crystallite coated
with 30 nm Au. The white dashed lines mark the YBCO crystallite
edges, exposing (100) facets. (b) Tunneling spectra measured along
the blue arrow shown in (a). The gap sizes are denoted, from 6.7
meV (near the crystal edge) to 2 meV (the smallest detectable
gap).  Note the corresponding increase of the zero bias
conductance.}\label{fig2}
\end{figure}

A correlated topography/spectroscopy measurement manifesting the
evolution of the gap structure with the distance from the
crystallite \textit{a}-axis facet is presented in figure
\ref{fig2}. The topographic image in Fig.\ref{fig2}(a) is of a 30
nm thick Au layer coating an YBCO film, showing a crystallite of
about 90 nm in size (marked by a white dashed line at the lower
right region). Tunneling spectra were obtained sequentially along
the line marked by the arrow, and are presented in
Fig.\ref{fig2}(b). It is evident that the gap size reduces with
increased distance from the crystallite edge, and at the same time
the zero bias conductance increases. As discussed in details in
Ref. \onlinecite{15}, such edges typically expose the (100) YBCO
surface. Note that the maximal observed gap, just at the edge -
6.7 meV, is still much smaller than the 20 meV gap of the bare
YBCO sample. This is consistent with the presence of the Au layer
that determines a minimal distance from the N-S interface,
corresponding to the Au film thickness (see Fig.\ref{fig1}(b)). At
larger distances from the edge, even for thinner gold layers, the
gap structure practically vanished (due to the increase of the
zero bias conductance it was hard to detect gap values much
smaller than 2 meV).

The results described above clearly indicate that the PE
originates primarily at the \textit{a}-axis YBCO surface, whereas
the contribution of the \textit{c}-axis surface is negligible, in
agreement with previous studies of the Josephson effect in related
systems.\cite{8}  This behavior reflects the quasi two-dimensional
\textit{d}-wave symmetry of the order parameter in YBCO.  However,
no indication of \textit{d}-wave symmetry was found for the
proximity-induced pair-amplitude in gold.  Unlike our previous
observations for YBCO films,\cite{15} the measured tunneling
spectra did not exhibit anisotropy when measured on different
faces of a gold grain.  In particular, zero bias conductance
peaks, which appear in tunneling spectra measured on the (110)
surface of pure YBCO, have never been observed on gold proximity
layers thicker than 7 nm (on the verge of full gold coverage).
They were observed only rarely for the thinnest layers that we
measured, 5 nm thick and less. It should be pointed out, however,
that (110) facets in the underlying YBCO films were not abundant
in our samples.  We therefore measured a (110) YBCO film
over-coated by a 7 nm Au film.  Here also, zero bias conductance
peaks were only rarely observed, and they were much weaker as
compared to those measured on the corresponding bare (110) YBCO
film. The proximity induced order parameter thus appears to be
predominantly \textit{s}-wave in nature, as will be further
discussed below.

\begin{figure}
\includegraphics[width=8cm]{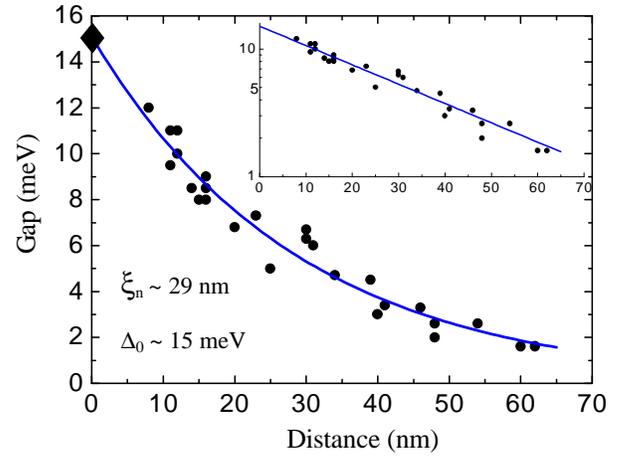}
\caption{Measured gap size as a function of distance from the N-S
interface (solid circles). The distance is estimated from the
distance to the crystallite edge, measured form the STM
topographic images, taking into account the nominal Au film
thickness. The solid line is a fit to a decaying exponential form,
with normal coherence length (penetration depth) $\xi_n \approx$
29 nm, and interface gap value $\Delta_0 \approx$ 15 meV (the
diamond symbol). The inset shows the data in a semi-log scale.
}\label{fig3}
\end{figure}

In figure \ref{fig3} we display an accumulation of the proximity
gaps measured on the gold surface as a function of the distance to
the closest (relevant) S-N interface (solid circles). This
distance was determined from the distance to the crystallite edge,
measured from the STM topographic images, taking into account the
nominal thickness of the gold film (that in many cases was larger
than the lateral distance to the crystallite edge).  The solid
line represents a best fit to the standard exponential decay form
of the gap,\cite{1,6} $\Delta(x) = \Delta_0exp(-x/\xi_n$).  This
fit was obtained for a penetration depth (normal coherence length)
of $\xi_n \approx$ 29 nm, and an interface gap value  $\Delta_0
\approx$ 15 meV. Note that the value of  n is much larger than the
rms roughness of the gold films (less than 1.5 nm), thus local
height fluctuations of the Au layer do not affect much the
measured gaps.

The value extracted for the penetration depth is very close to
that estimated for the dirty limit,\cite{1} $\xi_N = (\frac{\hbar
\mathit{V}_{FN} l_N}{6\pi k_B T}) \approx 33$ nm. The latter
estimation was obtained assuming that the elastic mean free path
$l_N$ in the gold film is governed by grain boundary scattering,
thus $l_N \sim 10$ nm, and taking the Fermi velocity in gold from
the literature, $V_{FN} = 1.4\times10^{6}$ m/s.

More interesting is the value we found for $\Delta_0$. This value
is similar to the magnitude of the \textit{s}-wave order parameter
component observed in YBCO in the point-contact experiments of
Kohen \textit{et al.},\cite{14} for the most transparent Au-YBCO
junctions measured.  This \textit{s}-wave component induced near
the surface of YBCO, which has a dominant $d_{x^{2}-y^{2}}$-wave
order parameter, was attributed to an anomalous N-S proximity
effect. We expect that in our samples, the Au-YBCO electrical
contact (or junction transparency) is at least as good as in the
most transparent junctions reported in Ref. \onlinecite{14} since
our Au films were deposited onto the YBCO surface in-situ, without
breaking the vacuum. Therefore, a strong \textit{s}-wave component
is probably induced in our case, and it is possible that it is
this component, and not the dominant $d_{x^{2}-y^{2}}$ order
parameter, which takes part in inducing the pair amplitude at the
N side of the Au-YBCO junctions.

In summary, tunneling spectroscopy correlated with topographic
measurements was used to observe the proximity effect on the
normal side of highly transparent YBCO-Au junctions. The
proximity-induced gap in Au decays exponentially with distance
from \textit{a}-axis YBCO facets, indicating that the PE is
primarily due to the (100) YBCO surface. Interestingly, while this
facet-selectivity reflects the anisotropic nature of
superconductivity in YBCO, the proximity-induced superconductivity
appeared to be isotropic, \textit{s}-wave in nature.  In
particular, the interface gap correlates well with the recently
observed proximity-induced \textit{s}-wave order parameter in
YBCO,\cite{14} and the coherence length in Au, $\xi_n \sim$30 nm,
conforms to `conventional' PE in the dirty-limit.

\acknowledgments {We thank Guy Deutscher for suggesting that the
most relevant length scale in our experiment is the distance from
an a-axis facet and not the Au thickness, and Amir Kohen for
helpful discussions. The research was supported in parts by the
Israel Science Foundation and the Heinrich Hertz Minerva center
for high-temperature superconductivity.}


\begin{references}
\bibitem {1}
G. {Deutscher} and P. G. {De Gennes}, {\emph{Superconductivity}}
(Marcel Dekker, Inc., New York, 1969).

\bibitem {2}
E. L. {Wolf}, {\emph{Principles of Electron Tunneling
Spectroscopy}} (Oxford University Press, New York, 1985).

\bibitem {3}
Y. {Levi}, O. {Millo}, N. D. {Rizzo}, D. E. {Prober} and L. R.
{Motowidlo}, Phys. Rev. B \textbf{58}, 15128 (1998).

\bibitem {4}
N. {Moussy}, H. {Courtois}, and B. {Pannetier}, Europhys. Lett.
\textbf{55}, 861 (2001).

\bibitem {5}
P. G. {de Gennes}, {\emph{Superconductivity of Metals and Alloys}}
(Benjamin, New York, 1966).

\bibitem {6}
W. {Belzig}, C. {Bruder}, and G. {Schon}, Phys. Rev. B
\textbf{54}, 9443 (1996).

\bibitem {7}
S. {Gueron}, H. {Pothier}, N. O. {Birge}, D. {Esteve} and M. H.
{Devoret}, Phys. Rev. Lett. \textbf{77}, 3025 (1996).

\bibitem {8}
M. A. M. {Gijs}, D. {Scholten}, Th. van {Rooy} and A. M.
{Gerrits}, Appl. Phys. Lett \textbf{57}, 2600 (1990).

\bibitem {9}
H. Z. {Durusoy}, D. {Lew}, L. {Lombardo}, A. {Kapitulnik}, T. H.
{Geballe} and M. R. {Beasley}, Physica C \textbf{266}, 253 (1996).

\bibitem {10}
E. {Polturak}, G. {Koren}, D. {Cohen}, E. {Aharoni} and G.
{Deutscher}, Phys. Rev. Lett. \textbf{67}, 3038 (1991).

\bibitem {11}
J. {Gao}, W. A. M. {Aarnink}, G. J. {Gerritsma} and H. {Rogalla},
Physica C \textbf{171}, 126 (1990).

\bibitem {12}
Y. {Ohashi}, J.  Phys. Soc. Jpn. \textbf{65}, 823 (1996).

\bibitem {13}
A. A. {Golubov} and M. Y. {Kpriyanov}, JETP Lett. \textbf{67}, 501
(1998).

\bibitem {14}
A. {Kohen}, G. {Leibovitch} and G. {Deutscher}, Phys. Rev. Lett.
\textbf{90}, 207005 (2003).

\bibitem {15}
A. {Sharoni}, G. {Koren} and O. {Millo}, Europhys. Lett.
\textbf{54}, 675 (2001).

\bibitem {16}
O. {Nesher}, G. {Koren}, E. {Polturak} and G. {Deutscher}, Appl.
Phys. Lett \textbf{72}, 1769 (1996).

\bibitem {17}
E. {Bar-Sadeh} and O. {Millo}, Phys. Rev. B \textbf{53}, 3482
(1996).

\end{references}
\end{document}